\documentclass{aa}
\usepackage{natbib,twoopt}
\bibpunct{(}{)}{;}{a}{}{,}  
\usepackage[varg]{txfonts}
\usepackage{graphicx}

\def\Ks{{K$_\mathrm{s}$}}

\def\mg{{$^\mathrm{m}$}}

\def\kms{{km sec$^\mathrm{-1}$}}
\def\km2s2{{km$^2$ sec$^\mathrm{-2}$}}
\def\kmskpc{{km sec$^\mathrm{-1}$ kpc$^\mathrm{-1}$}}

\def\rfr{{dF$_\mathrm{r}$}}
\def\lII{{l$^\mathrm{II}$}}
\def\bII{{b$^\mathrm{II}$}}
\def\lTe{{$\log(T_\mathrm{eff})$}}
\def\Te{{$T_\mathrm{eff}$}}
\def\aVr{{$\overline{V_r}$}}
\def\mVr{{$V_r^m$}}

\begin{document}

\title{The spiral potential of the Milky Way\thanks{Based on
    observations collected at the European Southern Observatory, Chile (ESO
    programme 097.B-00245, 099.B-0697)} \thanks{The catalog of
    radial velocities is only available in electronic form at the CDS via
    anonymous ftp to cdsarc.u-strasbg.fr (130.79.128.5) or via
    http://cdsweb.u-strasbg.fr/cgi-bin/qcat?J/A+A/}}

\subtitle{}
\titlerunning{The spiral potential of the Milky Way}
\author{P.~Grosb{\o}l\inst{\ref{inst1}} \and G.~Carraro\inst{\ref{inst2}} }
\offprints{P.~Grosb{\o}l}
\institute{European Southern Observatory,
  Karl-Schwarzschild-Str.~2, D-85748 Garching, Germany\\
  \email{pgrosbol@eso.org}\label{inst1}
\and
  Dipartimento di Fisica e Astronomia,
  Universita’ di Padova, Vicolo Osservatorio 3, I-35122 Padua, Italy\\
  \email{giovanni.carraro@unipd.it}\label{inst2}
}

\date{Received ??? / Accepted ???}

\abstract{The location of young sources in the Galaxy suggests a four-armed
  spiral structure, whereas tangential points of spiral arms observed in the
  integrated light at infrared and radio wavelengths indicate that only two
  arms are massive. }
{Variable extinction in the Galactic plane and high light-to-mass ratios of
  young sources make it difficult to judge the total mass  associated with the
  arms outlined by such tracers.  The current objective is to estimate the mass
  associated with the Sagittarius arm by means of the kinematics of the stars
  across it. }
{Spectra of 1726 candidate B- and A-type stars within 3\degr\ of the Galactic
  center (GC) were obtained with the FLAMES instrument at the VLT with a
  resolution of $\approx$6000 in the spectral range of 396--457\,nm. Radial
  velocities were derived by least-squares fits of the spectra to synthetic
  ones. The final sample was limited to 1507 stars with either Gaia DR2
  parallaxes or main-sequence B-type stars having reliable spectroscopic
  distances. }
{ The solar peculiar motion in the direction of the GC relative to the local
  standard of rest (LSR) was estimated to U$_\sun$ = 10.7$\pm$1.3\,\kms.  The
  variation in the median radial velocity relative to the LSR as a function of
  distance from the sun shows a gradual increase from slightly negative values
  near the sun to almost 5\,\kms\ at a distance of around 4\,kpc.  A
  sinusoidal function with an amplitude of $3.4\pm1.3$\,\kms\ and a maximum at
  $4.0\pm0.6$\,kpc inside the sun is the best fit to the data.  A positive
  median radial velocity relative to the LSR around 1.8\,kpc, the expected
  distance to the Sagittarius arm, can be excluded at a 99\% level of
  confidence.  A marginal peak detected at this distance may be associated with
  stellar streams in the star-forming regions, but it is too narrow to be
  associated with a major arm feature.  }
{A comparison with test-particle simulations in a fixed galactic potential
  with an imposed spiral pattern shows the best agreement with a two-armed
  spiral potential having the Scutum--Crux arm as the next major inner arm.
  A relative radial forcing \rfr$ \approx 1.5$\% and a pattern speed in the
  range of 20--30\,\kmskpc\ yield the best fit.  The lack of a positive velocity
  perturbation in the region around the Sagittarius arm excludes it from being
  a major arm.  Thus, the main spiral potential of the Galaxy is two-armed,
  while the Sagittarius arm is an inter-arm feature with only a small mass
  perturbation associated with it.  }
\maketitle
\section{Introduction}
The first evidence of spiral structure in the Milky Way was presented in the
1950s; it was obtained from optical observations of early-type stars first, and
from \ion{H}{I} surveys shortly thereafter
\citep[see][]{gingerich85,carraro15a}. Both
techniques have several well-known disadvantages, yet they were
extensively used in the past, and nowadays there are even claims that the
issue of the spiral structure of our Galaxy has been solved \citep{hou14}.
While optical observations are limited by the extinction problem, \ion{H}{I}
surveys suffer from the inability to derive reliable distances from radial
velocity, and from the fact that \ion{H}{I} is almost evenly distributed
across the Galactic disk \citep{liszt85}.  Traditionally, long-wavelength
observations depict a  two-armed Milky Way, while optical observations favor a
four-armed Milky Way.  Recent, extensive mapping of young objects  (e.g.,
OB-associations, \ion{H}{II}-regions, and young clusters) were presented by
\citet{russeil03} and revealed indications of a four-armed spiral structure.  On
the contrary, the tangent associated with the Sagittarius arm observed in
near-infrared bands shows no significant increase in integrated intensity
\citep{drimmel00}.  This suggests that little mass is associated with this arm,
as near-infrared surface brightness is well correlated to stellar mass
density, and that the Galaxy has a two-armed structure with Scutum--Crux as
the next major arm inside the sun.

These two scenarios need not contradict
each other as secondary shocks in the gas can be produced by a two-armed spiral
perturbation \citep{yanez08} leading to increased star formation between the
major arms.  A four-armed gas spiral may also be excited by the Galactic bar
\citep{englmaier99, englmaier06}.  For external spiral galaxies, the
appearance in the near-infrared is much smoother with weaker inter-arm
features than on visual images \citep{block94,gp98,eskridge02} as the former
emphasizes the cold stellar disk population and therefore the surface density
of the disks.

The structure of star-forming regions is important, but the shape of the
spiral potential is essential for the dynamics of the Galaxy.  The high
light-to-mass ratio of young objects makes it very difficult to deduce the
mass distribution from them.  A more direct method of estimating the mass
variation associated with nearby spiral arms is star counts of a well-defined
stellar population as a function of the distance from the sun.  This was done
for the Perseus arm in the anti-center direction using early-type stars for
which individual distances can be determined \citep{monguio15}.  For the
Sagittarius arm, the high and patchy extinction toward the Galactic center
(GC) makes it impossible to conduct a reliable star count.  The potential
perturbation can also be evaluated by measuring the velocity field of a
relaxed stellar population across arms.  One concern for very young stars is
that they may not be fully dynamically relaxed and still be biased by their
initial velocity.  An unbiased, random sample of velocities of a well-defined
stellar population allows an estimate of the perturbation independent of the
high, patchy extinction toward the GC assuming that they are uncorrelated.

The current paper studies the variations in radial velocities as a function of
distance for a sample of late B- and A-stars toward the GC with the aim of
determining a velocity perturbation associated with the Sagittarius arm.  The
sample is described in Sect.\,\ref{sec:obs} where  the observations are
 detailed as well.  The next section presents the reduction of the data and the
derivation of individual radial velocities and distances for the sources.
Possible models for the spiral potential in the Milky Way is discussed in
Sect.\,\ref{sec:dis}, while the conclusions are given in Sect. 5.

\begin{table}[t]
\caption[]{Summary of VLT/FLAMES observations.}
\label{tab:obs}
\centering
\begin{tabular}{lrrcccrr}
 \hline\hline
 Field &     
 \multicolumn{1}{c}{\lII} & \multicolumn{1}{c}{\bII} & MJD &
 n$_\mathrm{t}$ & n$_\mathrm{p}$  \\ \hline
  GCF-01 & -0\fdg09275 & -0\fdg34095 & 57529\fd41204 & 115 & 49 \\
  GCF-02 &  0\fdg06288 & -0\fdg90142 & 57566\fd31329 & 114 & 36 \\
  GCF-03 & -0\fdg30632 & -1\fdg16001 & 57622\fd14238 & 110 & 38 \\
  GCF-04 & -0\fdg79643 & -1\fdg41936 & 57622\fd18496 & 108 & 46 \\
  GCF-14 &  0\fdg51068 & -0\fdg81578 & 57633\fd11230 & 113 & 48 \\
  GCT-17 & -1\fdg22128 &  1\fdg46380 & 57882\fd17536 & 113 & 27 \\
  GCT-01 &  0\fdg78438 & -1\fdg27567 & 57889\fd25226 & 118 & 54 \\
  GCT-04 &  0\fdg22818 & -0\fdg62989 & 57889\fd29519 & 113 & 50 \\
  GCT-08 &  0\fdg27198 & -1\fdg58277 & 57889\fd33549 & 118 & 41 \\
  GCT-25 & -0\fdg50220 & -2\fdg22001 & 57889\fd37677 & 119 & 41 \\
  GCT-22 & -2\fdg71804 &  0\fdg79060 & 57890\fd21737 & 119 & 20 \\
  GCT-28 & -2\fdg29749 &  1\fdg18844 & 57890\fd25625 & 113 & 17 \\
  GCT-07 & -0\fdg17626 & -0\fdg61435 & 57890\fd29912 & 116 & 40 \\
  GCT-11 & -0\fdg14420 & -1\fdg77122 & 57890\fd33841 & 118 & 35 \\
  GCT-12 &  0\fdg74070 & -1\fdg65106 & 57890\fd37711 & 119 & 35 \\
    \end{tabular}
\end{table}

\section{Targets and observations}
\label{sec:obs}
Velocity variations across arms in grand-design spirals measured in \ion{H}{I}
are on the order of 10\,\kms\ (see, e.g., \citealt{visser80,visser80a} for
M81). This sets an upper limit for the expected velocity change of a stellar
population since its response to a potential perturbation decreases with
higher velocity dispersion \citep{lin69,shu70a,mark76}.  A velocity
variation with this amplitude is easier to detect in a population with an
intrinsic low-velocity dispersion,  such as stars with ages of less
than 2\,Gyr \citep{yu18}.
The population should also be old enough to be dynamically relaxed in the
Galactic potential that required several encounters with the spiral
perturbations, i.e.,  at least 300\,Myr.  Thus, late B- and A-stars are the best
populations for the study of the effects of the spiral potential on the
stellar velocity field.  Selecting targets towards the GC, we can avoid any
significant influence of the Galactic differential rotation on the measured
velocity variation, which outweighs other issues like crowding and high
extinction.

The velocity dispersion of B- and A-stars in the solar neighborhood is around
20\,\kms\ \citep{yu18} which indicates that a sample of several hundred
sources is required for a $5\sigma$ detection of an average velocity
perturbation of 5--10\,\kms.  A critical point for distinguishing between a two-
or four-armed spiral potential in the Milky Way is the velocity perturbation
associated with the Sagittarius arm, which is located 1.8\,kpc inside the solar
radius in the direction toward the GC \citep{reid14,wu14} corresponding to a
distance modulus of 11\mg.  A sample of early-type stars with
B$<$15\mg\ should include sources beyond the Sagittarius arm, even with visual
extinctions in the range of 5\mg.

Due to the high extinction toward the GC, it is difficult to select B- and
A-stars based on their colors since nearby late-type stars have colors similar
to highly reddened early-type stars.  Therefore, the prime targets were taken
from the catalog of early-type stars by \citet{grosbol16} based on objective
prism observations.  The positional errors of this catalog
(i.e., $\sigma_\alpha \approx 1\arcsec, \sigma_\delta \approx 6\arcsec$) did
not allow a direct cross-match with other catalogs, due to the high stellar
density toward the GC.  Selecting sources from VVV \citep{vvv} with the
reddening corrected color index $Q = (H-K) - 0.563\times(J-H)$
\citep{indebetouw05} matching early-type stars (i.e. $-0\fm1 <Q<0\fm1$) and
\Ks$<$14\mg, the density was reduced enough to secure a unique matching of the
targets.

The most efficient facility for observing spectra of the candidates was the
FLAMES instrument \citep{pasquini02} at the ESO/VLT, which in its
GIRAFFE/MEDUSA mode offers 130 fibers with a 1\farcs2 diameter in a circular
field with a radius of 15\arcmin.  A total of 15 fields were defined selecting
the area with the highest density of prime targets, yielding 17--54 sources per
field.  Additional candidates were selected for the remaining fibers by
picking VVV sources with reddening corrected colors corresponding to
early-type stars and estimated blue magnitudes B$<$15\mg.

\begin{figure}[t]
  \resizebox{\hsize}{!}{\includegraphics{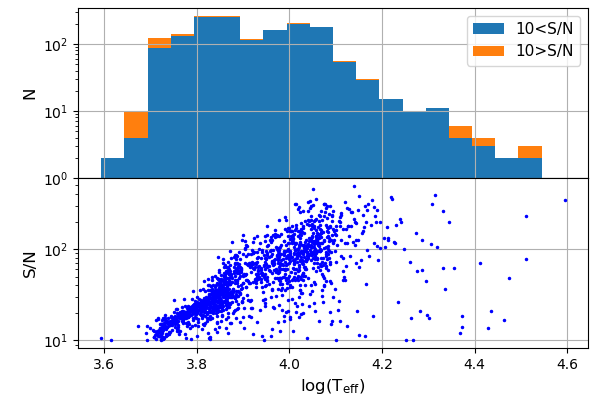}}
   \caption[]{Distribution of sources as a function of their effective
     temperature \Te.  The upper diagram gives the histogram of sources while
     the lower one shows the signal-to-noise ratio (S/N) both in logarithmic
     scale. }
  \label{fig:tsn}
\end{figure}

Five FLAMES/GIRAF fields were observed in 2016 (i.e., ESO P97) for which VVV
positions were used to center the fibers, leading to some errors due to the
differences in epoch.  The observations of the ten remaining fields were done in
2017 for which Gaia DR1 positions \citep{gaia16b,gaia16a,arenou17} were
adopted, which significantly improved the centering and thereby their
signal-to-noise ratio (S/N) for the same apparent magnitude. The LR2 grating
was used with a spectral resolution R = 6000 in the range of $\lambda$ =
396--456\,nm.

The observations are summarized in Table\,\ref{tab:obs}, which lists the 15
fields with their Galactic coordinates (\lII,\bII) and the Modified Julian
Date (MJD) of their mean exposure. The total number of fibers allocated,
n$_\mathrm{t}$, is also given including the fibers used for prime targets,
n$_\mathrm{p}$. For each field, 7--16 fibers were allocated to sky positions
with no stars visible on the Digital Sky Survey 2 images using Aladin
\citep{aladin}.  Four exposures of 660\,sec duration (in P97 only 600\,sec) 
were made of each field to allow for removal of cosmic ray events and extend
the dynamic range. The observations were conducted in service mode and yielded
a total of 1726 spectra of which 1505 had a S/N higher than 10. The seeing was
in the range of 0.8--1.4\arcsec, while the sky conditions were specified to
allow
for thin clouds.  A cross-matching based on positions gave 1635 sources with
\Ks-magnitudes in the VVV catalog, while 1608 were listed in the Gaia DR2
catalog with parallaxes \citep{gaia18a}.

\begin{figure}
 \resizebox{\hsize}{!}{\includegraphics{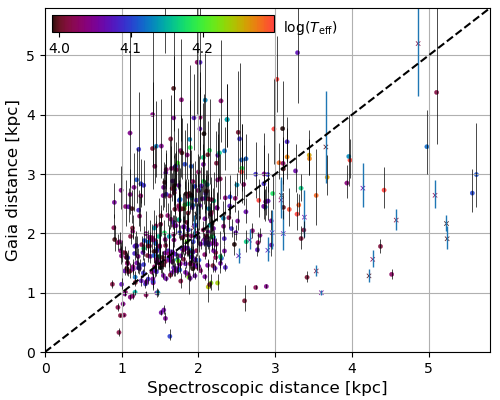}}
  \caption[]{Distances from Gaia DR2 parallaxes as a function of the distances
    determined from the spectra observed for stars with 4.0$<$\lTe.  Stars
    with $\log(g)<3.5$ are indicated by crosses.}
 \label{fig:rr}
\end{figure}

\section{Radial velocities and distances}
\label{sec:red}
The basic reductions including flat-fielding and wavelength calibration were
done via the standard ESO pipeline (Giraf/2.16.1) as processed by the
observatory.  For each field, the four exposures were averaged after outliers
(e.g., cosmic ray events) were removed with a median filter.  The spectra were
normalized to the continuum estimated by fitting a low-order polynomial to the
median values of sections with a few spectrum lines.  Radial velocities were
derived by matching the spectra with model grids of synthetic spectra
performing both a cross-correlation and least-squares fitting.  The synthetic
spectra were smoothed with a Gaussian filter and rebinned to match the
observed spectra.  In addition, a high-pass filter was applied to both
synthetic and
observed spectra to reduce biases due to errors in the normalization.  The
least-squares fits were preferred as they are less biased by the Balmer lines
than the cross-correlation estimates. The spectral range for the comparison
was limited to 399--454\,nm as the spectra did not fully include the
H$_\epsilon$ line.  In addition to the fit to this range, five subsections of the
spectra were used including two centered on the Balmer lines and three in the
regions in between.  This allowed us to verify whether the velocities
estimated from
the Balmer lines and metal lines were consistent.  The weighted mean of these
five estimates was adopted as the radial velocity after which the barycentric
correction was added.  Several grids were applied, such as the POLLUX
\citep{palacios10}, UVBLUE \citep{rodriguez05}, and PHOENIX databases
\citep{husser13}.  Since no significant differences between these grids were
found, the UVBLUE database was used as it provides a finer mesh.  Due to the
metallicity gradient in the Galactic disk, it is expected that stars inside
the solar radius have higher metallicities than the sun.  Grids with [M/H]=0.3
were adopted since models with higher metallicities gave worse fits.  In
addition to radial velocity, effective temperature \Te\ and surface gravity
$g$ for the individual stars were estimated through the least-squares fitting.
The distribution of sources is shown in Fig.~\ref{fig:tsn} as a function of
their effective temperature \Te.  The surface gravity could only be determined
reliably for stars with \lTe$>$4.0 for which Balmer and \ion{He}{I} lines are
sensitive to $g$.  Typical errors for the radial velocities are 3\,\kms\ for
stars with \lTe$>$3.9, while  for cooler stars they increase to around 5\,\kms.

Since many sources were selected based on their near-infrared colors, the
sample includes more than 100 late-type stars (see Fig.~\ref{fig:tsn}) with
\Te\ in the range in which the Gaia DR2 database \citep{gaia18a} provides
radial velocities.  A total of 63 common sources with both Gaia and FLAMES
radial velocities were identified.  A linear regression gave $V_r^{Gaia} =
2.5\pm1.4 + 0.997\pm0.038 \times V_r$.  The velocities measured were corrected
for this offset by adding 2.5\,\kms.

\begin{figure}
  \resizebox{\hsize}{!}{\includegraphics{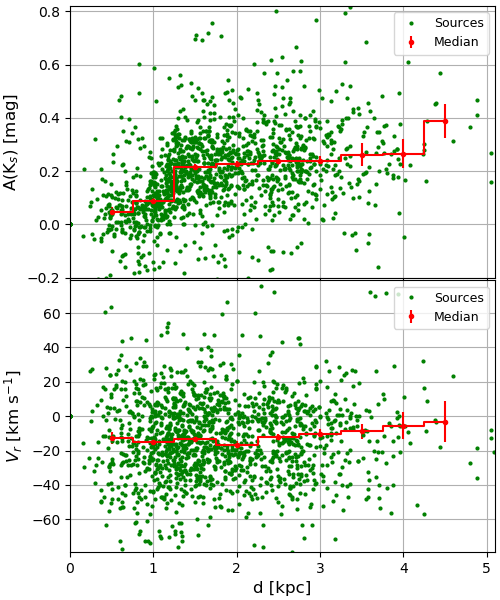}}
  \caption[]{Near-infrared extinction A(\Ks) and barycentric radial velocity
    $V_r$ as a function of distance $d$ from the Sun.}
  \label{fig:rav}
\end{figure}

Absolute magnitudes and intrinsic colors for the sources were obtained from
the Padova evolutionary tracks \citep{bressan12,marigo17} using the \Te\ and
$g$ values derived.  Models with  metallicity Z=0.03 were adopted.  The VISTA
filters were used as the best proxy for the VVV color system.  The color index
(Z-H) was used to obtain the largest wavelength difference of the bands while
omitting \Ks,\ which may be affected by Br$_\gamma$ emission.  Extinctions were
derived from the intrinsic colors using an extinction law $\lambda^{-\beta}$
with $\beta=1.517$ \citep{schlafly11} yielding the excess E(Z-H) = 0.302\,A(V)
= 2.364\,A(\Ks).  Distances were calculated from the \Ks-band colors applying
the A(\Ks) extinction correction when positive. The 1\arcsec\  aperture VVV 
magnitudes were used in order to reduce the contamination from nearby stars in
the very crowded fields.

Although only five sources show clear double line profiles in their
Balmer lines,
we must assume that a large fraction of the sources are multiple-star systems
\citep{duchene13}.  In the worst-case scenario where all systems are binaries
with equal luminosity, the spectroscopic distances should be increased by 40\%.
In a more realistic case, the correct distances would be 10--20\% longer and
potential features in the velocities as a function of distance would be
smoothed. 

The near-infrared extinction A(\Ks) as a function of distance from the sun,
$d$, is shown in Fig.~\ref{fig:rav} where  the median values in radial
bins of 0.5\,kpc are also indicated with corresponding error bars.
The extinction
displays a large scatter, but its median increases to a distance of nearly
1.4\,kpc, after which it remains nearly constant.  The flat behavior at larger
distances is likely a selection effect as only sources with a low extinction
will appear in the current magnitude-limited sample. A second rise in the
extinction at a distance of 4.3\,kpc is suggested.  Both of these step-wise
increases may be caused by the star-forming regions in the arms.

\begin{table}[t]
  \caption[]{List of the number of sources N, mean velocity \aVr, standard
    deviation $\sigma(V_r)$, and median velocity \mVr\ in \kms\ including their
    errors in distance bins centered on $d$ (kpc).  The first section provides
    the values for equal radial bins, while the second section  for equal
    number of sources per bin.}
\label{tab:vr}
\centering
\begin{tabular}{crrcr}
 \hline\hline
 $<$$d$$>$ & \multicolumn{1}{c}{N} & \multicolumn{1}{c}{\aVr} &
   \multicolumn{1}{c}{$\sigma(V_r)$} & \multicolumn{1}{c}{\mVr} \\ \hline
   0.55 &  119 &  -13.2\,$\pm$\,2.7 &  29.3 &  -12.7\,$\pm$\,3.4 \\
   1.03 &  313 &  -15.6\,$\pm$\,1.6 &  28.5 &  -14.8\,$\pm$\,2.0 \\
   1.48 &  373 &  -13.3\,$\pm$\,1.2 &  23.9 &  -13.0\,$\pm$\,1.6 \\
   1.98 &  240 &  -15.2\,$\pm$\,1.6 &  24.9 &  -16.5\,$\pm$\,2.0 \\
   2.51 &  220 &  -12.1\,$\pm$\,1.6 &  23.9 &  -12.3\,$\pm$\,2.0 \\
   2.97 &  133 &  -13.5\,$\pm$\,2.0 &  23.1 &  -10.1\,$\pm$\,2.5 \\
   3.47 &   64 &  -11.6\,$\pm$\,3.5 &  28.2 &   -8.7\,$\pm$\,4.4 \\
   3.93 &   29 &   -6.6\,$\pm$\,6.4 &  34.3 &   -5.5\,$\pm$\,8.0 \\
   4.75 &   16 &  -12.2\,$\pm$\,5.0 &  19.9 &  -10.6\,$\pm$\,6.2 \\
\hline\hline
   0.52 &  100 &  -14.3\,$\pm$\,3.1 &  30.7 &  -14.2\,$\pm$\,3.8 \\
   0.81 &  100 &  -14.0\,$\pm$\,3.0 &  30.3 &  -14.7\,$\pm$\,3.8 \\
   1.00 &  100 &  -19.6\,$\pm$\,2.6 &  25.6 &  -15.8\,$\pm$\,3.2 \\
   1.15 &  100 &  -14.4\,$\pm$\,2.7 &  27.3 &  -13.9\,$\pm$\,3.4 \\
   1.27 &  100 &  -14.6\,$\pm$\,2.6 &  25.8 &  -13.0\,$\pm$\,3.2 \\
   1.38 &  100 &  -15.2\,$\pm$\,2.5 &  25.2 &  -13.6\,$\pm$\,3.2 \\
   1.52 &  100 &  -13.4\,$\pm$\,2.4 &  23.8 &  -14.8\,$\pm$\,3.0 \\
   1.67 &  100 &   -7.8\,$\pm$\,2.3 &  23.1 &   -5.5\,$\pm$\,2.9 \\
   1.84 &  100 &  -13.8\,$\pm$\,2.8 &  27.7 &  -13.8\,$\pm$\,3.5 \\
   2.02 &  100 &  -15.4\,$\pm$\,2.0 &  20.0 &  -18.2\,$\pm$\,2.5 \\
   2.27 &  100 &  -13.0\,$\pm$\,2.7 &  27.3 &  -10.0\,$\pm$\,3.4 \\
   2.50 &  100 &  -11.0\,$\pm$\,2.3 &  22.6 &  -12.4\,$\pm$\,2.8 \\
   2.72 &  100 &  -16.9\,$\pm$\,2.4 &  24.0 &  -16.3\,$\pm$\,3.0 \\
   3.03 &  100 &  -12.1\,$\pm$\,2.2 &  21.6 &  -10.1\,$\pm$\,2.7 \\
   3.70 &  100 &  -10.6\,$\pm$\,3.0 &  29.7 &   -8.3\,$\pm$\,3.7 \\
\end{tabular}
\end{table}

Of the 1608 sources with Gaia parallaxes, 469 stars have reliable $\log(g)$
estimates (i.e., \lTe$>4.0$) and therefore spectroscopic distances.  These
distances are, in general, consistent with the parallaxes (see
Fig.~\ref{fig:rr}) except for two groups of stars.  One group with around
60 stars
has significantly shorter distances than given by the Gaia parallaxes, likely
caused by multiple-star systems.  This group has on average larger errors
which may be due to their multiplicity.  Another set of almost 80 stars has
much larger spectroscopic distances as the spectral fits suggested
that they were 
giants, i.e., $\log(g)<3.5$.  Distances calculated from the Gaia DR2 parallaxes
were adopted for 1484 sources with parallaxes larger than 3.0 times their
error.  The error distribution of distances is skewed to longer distances
assuming Gaussian errors for the parallaxes.  This effect was estimated using
a Monte Carlo simulation of errors which showed that the shift is less than
0.5\,kpc for distances up to 3\,kpc.  Thus, in the worse-case scenario where
all parallaxes have 33\% relative errors, the distances will be underestimated
by around 20\%.  In addition, spectroscopic distance were used for 45
main-sequence stars with \lTe$>4.0$ and S/N$>$10 yielding a total sample of
1529 sources.  It should be noted that distances are correlated with the
\Te\ estimates and therefore ages since the sample is magnitude limited.

The lower panel of Fig.~\ref{fig:rav} displays the barycentric radial
velocities together with their median $V^m_r$ as a function of their distance
$d$, while the numeric values are listed in Table~\ref{tab:vr}.  The velocity
distribution contains 12 outliers with velocities $|V_r-$\mVr$| >
3\times\sigma(V_r^m) = 110$\,\kms, all with positive residuals while 8 had
S/N<10.  They were omitted as they are unlikely to be associated with the young
stellar disk population.  This left 1517 stars with a median velocity of
-13.2\,$\pm$0.7\,\kms\ and a dispersion of 26.8\,\kms.  The former is
consistent with the peculiar motion of the sun relative to LSR U$_\sun$ =
11.1\,$\pm$0.7\,\kms\ \citep{schonrich10}.  The dispersion is slightly higher
than that measured for the local young population \citep{yu18} due to the
variation in the average as a function of distance and multiple-star systems.
The mean radial velocities corrected to the LSR are negative close to the sun
and then increase to positive values with a maximum at around 4\,kpc.  The
peak-to-peak velocity variation is almost 8\,\kms.

\begin{figure}
  \resizebox{\hsize}{!}{\includegraphics{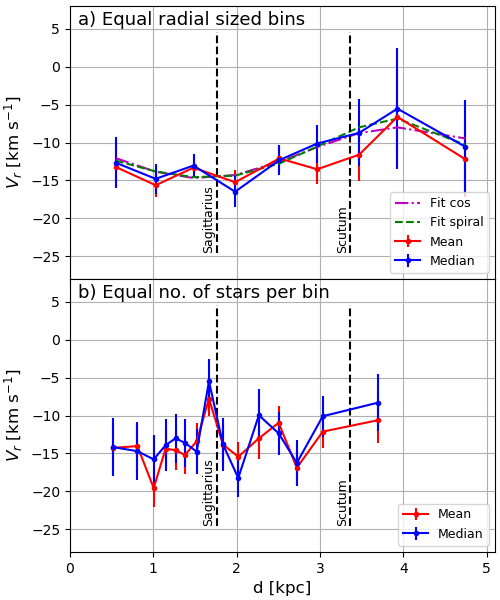}}
  \caption[]{Mean and median radial velocities relative to the sun as a
    function of distance $d$ from the sun binned {\bf a)} in 0.5\,kpc bins,
    and {\bf b)} in bins with 100 sources in each. Vertical lines indicate the
    locations of masers in the Sagittarius and Scutum arms.}
  \label{fig:rvf}
\end{figure}

\begin{table}[t]
  \caption[]{Parameters for the spiral potential models.}
\label{tab:mdl}
\centering
\begin{tabular}{lcrrrrrr}
 \hline\hline
 \multicolumn{1}{c}{Model} & m & \multicolumn{1}{c}{$i$} &
 \multicolumn{1}{c}{$r_a$} & \multicolumn{1}{c}{$A_4/A_2$} &
 \multicolumn{1}{c}{$\Omega_b$} & \multicolumn{1}{c}{$h_b$} &
 \multicolumn{1}{c}{$n_b$}\\ \hline
    {\tt s2ann}  &  2   & -12\fdg3 &  5.0 & 0.00 &   - &   -  &  -  \\
    {\tt s2arr}  &  2   & -12\fdg3 &  5.0 & 0.00 &  30 &  4.0 &  4  \\
    {\tt s2arf}  &  2   & -12\fdg3 &  5.0 & 0.00 &  40 &  4.0 &  4  \\
    {\tt s2bnn}  &  2   &  -7\fdg4 &  6.6 & 0.00 &   - &   -  &  -  \\
    {\tt s2cnn}  &  2-4 & -12\fdg3 &  5.0 & 0.25 &   - &   -  &  -  \\
    {\tt s2dnn}  &  2-4 & -12\fdg3 &  5.0 & 0.50 &   - &   -  &  -  \\
    {\tt s2enn}  &  2-4 & -12\fdg3 &  5.0 & 0.75 &   - &   -  &  -  \\
    {\tt s4ann}  &  4   & -14\fdg5 &  6.6 &  -   &   - &   -  &  -  \\
    {\tt s4arr}  &  4   & -14\fdg5 &  6.6 &  -   &  30 &  4.0 &  4  \\
    {\tt s4arf}  &  4   & -14\fdg5 &  6.6 &  -   &  40 &  4.0 &  4  \\
\end{tabular}
\end{table}

\section{Discussion}
\label{sec:dis}
The statistics of the velocities were computed in 0.5\,kpc radial bins in the
range of 0.5--5.0\,kpc containing a total of 1507 stars; they are listed in
Table~\ref{tab:vr}, where the two last bins were joined due to the small number
of sources.  The variation looks like a sinusoidal function (see
Fig.~\ref{fig:rvf}a), as expected for a density wave perturbation
\citep{lin64}, with a maximum of around 4\,kpc.  This trend is consistent with
that measured for older stars using RAVE and Gaia DR2 data
\citep{siebert12,carrillo18,gaia18d} in the overlapping range below a distance
of 1.5\,kpc in the plane. In a density wave scenario with Perseus and
Sagittarius arms as major arms, the sun would lie in the inter-arm region with
the LSR having a lower values of its U component than that measured close to
the Sagittarius arm.  There is no indication of a significant velocity peak
close to the expected distance of the Sagittarius arm (i.e., 1.8\,kpc).  The
mean velocity of the 608 stars within 0.5\,kpc of the arm is
-14.5$\pm$1.2\,kms,\ which is less than U$_\sun$
=11.1$\pm0.7$\,\kms\ \citep{schonrich10} at a 99\% level of confidence.

A weighted fit of the median values of the radial velocities toward the GC
with the function $V_r(d) = A_v\cos(2\pi(d-D_o)/l_v) + V_o$ yields an
amplitude $A_v = 3.4\pm1.3$\,\kms, a maximum close to $D_o = 4.0\pm0.6$\,kpc,
a wavelength $l_v = 4.8\pm1.3$\,kpc, and a velocity offset $V_o =
-11.4\pm1.2$\kms.  Using a function for a logarithmic spiral $V_r(d) =
A_v\cos(\ln(((R_\sun-d)/R_o)/t_v) + V_o$, we obtain $A_v = 3.9\pm1.4$\,\kms,
a maximum at $\ln(R_o/\mathrm{kpc}) = 1.48\pm0.08$ with a scale $t_v =
0.13\pm0.02,$ and an offset $V_o = -10.7\pm1.2$\,\kms.   Amplitude and
phase of the velocity variation are well determined; however,
the wavelengths $l_v$ and
$t_v$ depend critically on the last bins and are therefore less reliable.
Furthermore, the wavelengths are likely to be underestimated due to the skewed
error distribution of distances.  The relative smooth variation of the median
values as a function of radius suggests that the mass perturbation must have
existed for a significant time as it otherwise could not have created such a
regular velocity response.  On the other hand, a more transient perturbation
\citep{grand15,baba18} cannot be excluded on the basis of the current data
which only cover a small area of the Galactic plane.

\begin{figure}
  \resizebox{\hsize}{!}{\includegraphics{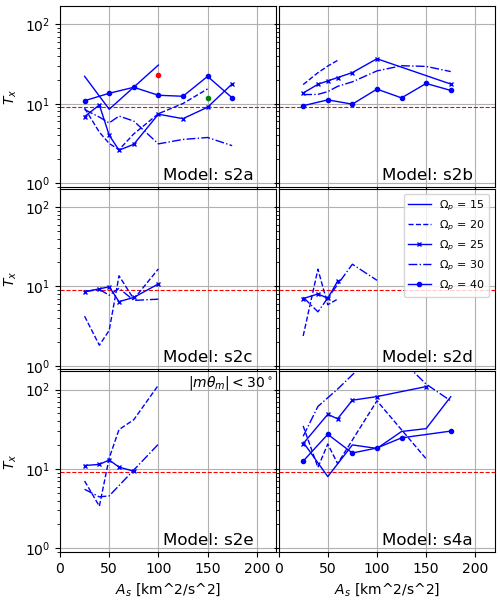}}
  \caption[]{Test variable $T_x$ for models with phases
    $|m\theta_m|<30\degr$ for the six models as a function of amplitude and
    pattern speed $\Omega_p$ in \kmskpc. The horizontal dashed line indicates
    the $T_x$ level for a model without spiral perturbations.}
  \label{fig:dm1}
\end{figure}

To increase the radial resolution near the sun where the sources density is
higher, bins with an equal number of sources were also used, as shown in
Fig.~\ref{fig:rvf}b.  This shows a general increase in the median velocity as
a function of distance similar to that seen with fixed radial bins, but also
two peaks at radii 1.7\,kpc and 2.3\,kpc.  They are only marginally significant
(i.e., at a 2--3$\sigma$ level) and are reduced both if smaller or
larger numbers
of sources per bin are chosen.  If they are real, they may be associated with
minor mass concentrations in the vicinity of the Sagittarius arm.  Furthermore,
moving groups or stellar streams could be the origin since the stars have
similar ages due to the distance--age correlation.  The short distance between
the peaks and their narrowness make it very unlikely that they are
associated with a global density wave feature in the Galactic potential
(e.g., a four-armed
density wave is expected to have an inter-arm distance of at least 3\,kpc).

\begin{figure*}
  \resizebox{\hsize}{!}{\includegraphics{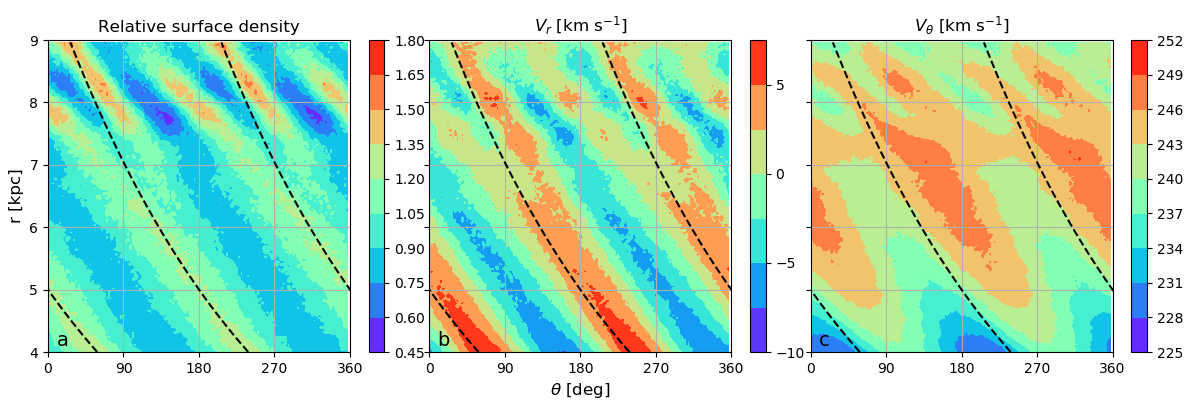}}
  \caption[]{Maps of the best test-particle simulation, {\tt s2c}, with
    $\Omega_p=20$\,\kmskpc\ and $A_2 = 40$\,\km2s2: a) relative surface
    density (normalized to unity in azimuth), b) $V_r = -v$ (i.e., positive for
    motion toward GC), and c) $V_{\theta}$.  The imposed spiral potential
    minima are indicated by dashed lines}
  \label{fig:tpm}
\end{figure*}

A simple fit of an analytic function (e.g., a sinusoidal) does not account for
possible nonlinear dynamical effects such as resonances. To include such
nonlinear effects a grid of test-particle simulations was computed in a
fixed axisymmetric potential with an imposed, density wave-like, spiral
perturbation \citep{lin64} (see Appendix~\ref{apx:pot} for details).
Three spiral
patterns were selected to agree with the observed locations of young sources
in the Perseus arm at 9.9\,kpc, the Sagittarius arm at 6.6\,kpc, or the Scutum
arm at 5.0\,kpc \citep{reid14,wu14}.  Two two-armed configurations were
considered,  one with arms corresponding to Perseus and Scutum with a
pitch angle $i=-12\fdg3$ (i.e., {\tt s2a}) and the other using Perseus and
Sagittarius as the major arms with $i=-7\fdg4$ (i.e., {\tt s2b}).
For the four-armed model, a pattern going through the Perseus and
Sagittarius arms with $i=-14\fdg5$
was selected (i.e., {\tt s4a}).  In addition, models with a bar perturbation
were computed with a bar mass equal to 10\% of the bulge mass and a scale
length of 4\,kpc.  The parameters for the models are summarized in
Table~\ref{tab:mdl} where $m$ is the number of arms, $r_a$ the location of the
next inner arm, $A_m$ the amplitude of the $m$-armed spiral, $\Omega_b$ the
bar pattern speed, while $h_b$ and $n_b$ are the bar shape parameters.  Three
two-armed models with a weaker $m=4$ component were also considered to simulate
the case of a higher harmonic response (i.e., sharper arms).  Pattern speeds
for the spirals were varied in the range 15--40\,\kmskpc,\ while amplitudes up
to 200\,\km2s2\ were applied corresponding to a radial force perturbation
relative to the axisymmetric field of \rfr $\approx$6\%.  The models were
integrated to a total time of 1 and 2\,Gyr to ensure that an equilibrium was
reached.

The radial velocity of the simulations were calculated in a polar coordinate
system centered on the GC and with the phase $\theta$ increasing
counterclockwise
with zero in the direction toward the sun.  To compare the simulations with
the radial velocities measured, they were rebinned to match the data binned in
equal radial bins (see first part of Table~\ref{tab:vr}) and 2\fdg8 in
$\theta$.  The test variable $T_x(\theta) =
\sum_{n=1}^{N_{bin}}{([V_x(n,\theta)-V^m(n)-V_o]/\sigma(V^m))^2}$ was computed
where $n$ is the radial bin, $N_{bin}$ the number of bins, and $x$ denotes the
test-particle simulation including its amplitude and pattern speed.  The value
of $T_x$ follows a $\chi^2$ distribution with $N_{bin}-2$ degrees of freedom
since the velocity offset $V_o$ was estimated.

\begin{figure}
  \resizebox{\hsize}{!}{\includegraphics{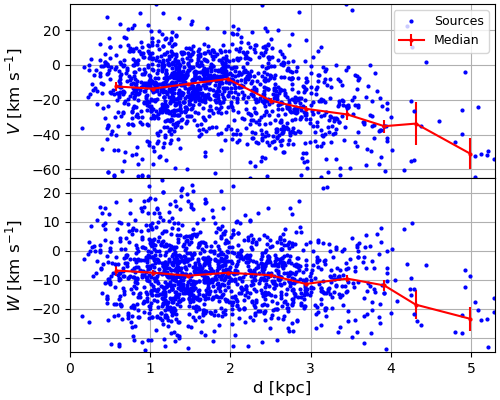}}
  \caption[]{Distribution of the velocity components V and W as a function of
    distance $d$ from the sun.}
  \label{fig:rvw}
\end{figure}

Acceptable models should have $T_x < 9.12$, which is the level estimated for
models without spiral perturbations.  Furthermore, the phase $\theta_m$ of the
minimum $T_x$ has to be close to zero in order to fit the distances observed
for the Sagittarius or Scutum arm. The current distances to the arms 
refer to star-forming regions (e.g., masers), which may have an offset
relative to the spiral potential minima.  A phase offset of up to 30\degr,
calculating 360\degr\ between arms, is possible if the star formation is
concentrated close to a shock in the gas flowing through the spiral
perturbation \citep{roberts69a,yuan81,gittins04}.  Thus, models with
$|m\theta_m| > 30$\degr\ should be rejected since they would not agree with
the observed positions of star-forming regions in the arms defined by maser
sources \citep{reid14}.  Finally, the velocity offset $V_o$ should be
consistent with the solar peculiar motion of 11.1\,\kms\ \citep{schonrich10}
within 3$\sigma$.  The minimum $T_x(|m\theta_p|<30\degr)$ values for models
integrated to 1\,Gyr are shown in Fig.~\ref{fig:dm1} as a function of
perturbation amplitude and pattern speed where the level of pure axisymmetric
models is indicated by a dashed line.  The actual values of $T_x$ are listed
in Table~\ref{tab:ps2a} together with the velocity offset $V_o$ and the phase
of the minimum $\theta_m$.  The models integrated to 2\,Gyr show very similar
results.

The best models with these three criteria are two-armed spiral potentials with
the Scutum arm as the next major, inner arm.  Minimum $T_x$ values are found
for pattern speeds $\Omega_p \approx$ 20--30\,\kmskpc\ and amplitudes $A_2
\approx 50$\,\km2s2, which corresponds to \rfr $\approx$1.5\%.  Two-armed
models with an additional m=4 harmonics (i.e., {\tt s2c}--{\tt s2e}) have
slightly smaller $T_x$ values.  This corresponds to a slightly sharper spiral
potential in azimuth.  The amplitude of this m=4 component depends on the
steepness of the velocity profile near 4\,kpc which is uncertain due to the
small number of stars.  Maps of relative surface density, radial velocity, and
azimuthal velocity for the best model, {\tt s2c}, with $\Omega_p =
20$\,\kmskpc\ and $A_2 = 40$\,\km2s2\ are shown in Fig.~\ref{fig:tpm}. This
pattern speed places the 4:1 resonance region close to the solar radius which
may explain some of the features seen in the velocity field in the solar
neighborhood \citep{gaia18d,kawata18,antoja18}.

Although a few models with the Sagittarius as the next major arm
(e.g., {\tt s2b}
and {\tt s4a}) have $T_x$ slightly smaller than a axisymmetric model, most of
these models have $|m\theta_m|>30\degr$ and can therefore be excluded.  Models
with an additional bar potential did not improve the match significantly as
the main effect was a small shift of the radial velocities with no radial
modulation except for resonance effects.

With Gaia DR2 proper motions, the tangential velocities of 1498 sources in the
sample can be calculated using \citet{johnson87} as seen in
Fig.~\ref{fig:rvw}.  The velocity component V in the direction of the Galactic
rotation shows a flat distribution out to 2\,kpc with $<$V$>$ =
-12.1\,$\pm$0.8\,\kms. At greater distances, a decline is seen due to the
differential rotation of the Galaxy.  A small peak near 2\,kpc is consistent
with the radial velocity variation, assuming a density wave perturbation (see
Fig.~\ref{fig:tpm}c); however, the uncertainty on the Galactic rotation curve
does not allow us to draw any conclusions.  Similar variations are seen by
\citet{gaia18d} and \citet{kawata18}, also based on the Gaia DR2.  The W
component perpendicular to the Galactic plane also displayed a flat velocity
distribution with $<$W$>$ = 6.8$\pm$0.5\,\kms\ to at least 2\,kpc, after
which a slow decline is observed.  The lack of a significant variation in W
makes it unlikely that the radial velocity changes observed are caused by an
external source (i.e., a recent encounter with a dwarf galaxy) since such a
perturber  would also leave a trace in W velocities.  The latter is likely an
effect of the spatial distribution of the stars which are mainly located above
the plane at large distances.  The average values of all three velocity
components
are consistent with the solar peculiar motion relative to the LSR as
determined by \citet{schonrich10}.

\begin{figure}
  \resizebox{\hsize}{!}{\includegraphics{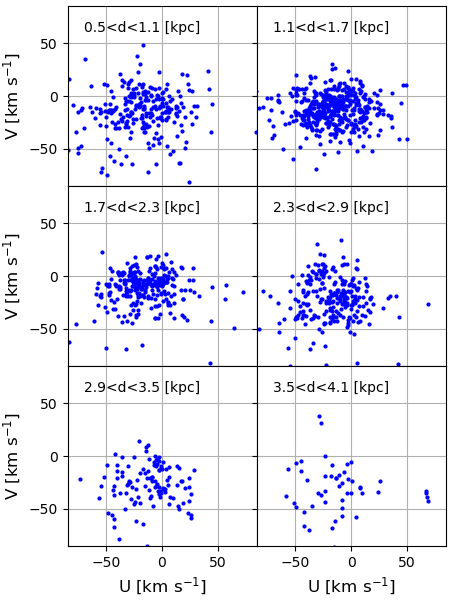}}
  \caption[]{U-V velocity ellipsoid as a function of distance $d$ from the sun.}
  \label{fig:uvd}
\end{figure}

The U-V velocity distributions are shown in Fig.~\ref{fig:uvd} grouped in
radial bins.  The closest bins display a marginal ellipsoidal distribution
with angle offset of around 30\degr\ from the direction toward the GC, whereas
the distributions at distances larger than 2.3\,kpc become more circular.
This may partly be due to the increasing errors in V as a function of
distance.  The angular offset is likely caused by the spiral perturbation
which suggests that it is dynamically important to at least a distance of
3\,kpc from the sun.  Closer to the GC, interaction with the bar potential
makes predictions uncertain.  Some substructures can be seen in several bins
(e.g., around 1.3 and 2.6\,kpc). They may be caused by stellar streams or
resonances, but the statistics are not sufficient for a detailed analysis.

\section{Conclusion}
\label{sec:con}
The radial velocities of 1726 stars within 3\degr\ of the GC were measured
with the FLAMES instrument at the VLT.  The final sample consisted of 1507
sources with reliable velocities and distances (i.e., either from Gaia DR2
or main-sequence B-stars). The variation in the median radial velocities
relative to the LSR as a function of distance from the sun shows a slow change
from slightly negative values to a maximum close to 4\,kpc.  A least-squares
fit of a sinusoidal function yields a velocity amplitude of
3.4$\pm$1.3\,\kms\ with a maximum at $d = 4.0\pm0.6$\,kpc and a wavelength
exceeding 4\,kpc.  This is consistent with the Scutum--Crux arm being the
next major arm inside the sun. It should be noted that the velocity amplitude is
likely to be underestimated as the average position of the stars is 40\,pc
 from the plane. The larger number of sources close to the sun allowed a
higher radial resolution which shows marginal velocity peaks at distances of
1.7 and 2.3\,kpc.  Although these peaks are in the range of the Sagittarius
arm, their narrowness suggests that they are  stellar streams near star-forming
regions rather than  global density wave-like perturbations.  The mean radial
velocity within 0.5\,kpc of the arm is less than that of the LSR at a 99\%
level of confidence, which excludes it as a major, long-lived mass
perturbation.

A fit of an analytic function to the velocity distribution does not account
for possible nonlinear effects (e.g., resonances) which even at a
relatively small
amplitude may play a role.  A set of test-particle simulations were computed
in a fixed axisymmetric potential with imposed spiral and bar perturbations
representing spiral patterns with either the Sagittarius or Scutum arm as the
next major inner arm.  The best agreement between the data and these models
was found for a two-armed pattern having the Perseus and Scutum arms as major
arms.  Pattern speeds in the range of 20--30\,\kmskpc\ and amplitudes around
50\,\km2s2\ (i.e., \rfr $\approx$ 1.5\%) were favored.  Two-armed models with a
sharper azimuthal perturbation (i.e., with an additional m=4 harmonic term)
gave slightly better fits, but depend on the shape of the velocity variation
measured near 4\,kpc, which is uncertain due to the small number of stars.  The
lifetime of the perturbation cannot be determined directly from the current
data.

The current data suggest that the spiral potential of the Milky Way is two-armed
with the Perseus and Scutum--Crux arms as majors.  With a velocity amplitude of
3.4\,\kms\ corresponding to \rfr$\approx$1.5\%, the perturbation is weak and
in the linear domain \citep{grosbol93}.  Marginal velocity peaks near the
Sagittarius arm may be associated with star formation in the arm, but are too
narrow to originate from a global perturbation.  This favors a view of the
Sagittarius arm as a weaker inter-arm feature with star formation excited
either by the bar \citep{englmaier99,englmaier06} or a secondary shock
\citep{yanez08}.
 
\begin{acknowledgements}
  We would like to acknowledge the detailed comments of the referee, which
  helped improve the paper.  It is also a pleasure to thank Dr. V. Korchagin
  for useful and stimulating discussions.  The ESO-MIDAS system and Python
  scripts with the {\it SciPy} package were used for the reduction and
  analysis of the data.  This work has made use of data from the European
  Space Agency (ESA) mission {\it Gaia}
  (\url{https:\\www.cosmos.esa.int/gaia}), processed by the {\it Gaia} Data
  Processing and Analysis Consortium (DPAC,
  \url{https://www.cosmos.esa.int/web/gaia/dpac/consortium}). Funding for the
  DPAC has been provided by national institutions, in particular the
  institutions participating in the {\it Gaia} Multilateral Agreement.
\end{acknowledgements}
\bibliographystyle{aa}
\bibliography{AstronRef}

\begin{thebibliography}{52}
\expandafter\ifx\csname natexlab\endcsname\relax\def\natexlab#1{#1}\fi

\bibitem[{Antoja {et~al.}(2018)Antoja, Helmi, Romero-G{\'o}omez, Katz,
  Babusiaux, Drimmel, {et~al.}}]{antoja18}
Antoja, T., Helmi, A., Romero-G{\'o}omez, M., {et~al.} 2018, arXiv, 1804.10196

\bibitem[{Arenou {et~al.}(2017)Arenou, Luri, Babusiaux, Fabricius, Helmi,
  Robin, {et~al.}}]{arenou17}
Arenou, F., Luri, X., Babusiaux, C., {et~al.} 2017, A\&A, 599, A50

\bibitem[{Baba {et~al.}(2018)Baba, Kawata, Matsunaga, Grand, \& Hunt}]{baba18}
Baba, J., Kawata, D., Matsunaga, N., Grand, R. J.~J., \& Hunt, J. A.~S. 2018,
  ApJ, 853, L23

\bibitem[{Bland-Hawthorn \& Gerhard(2016)}]{bland16}
Bland-Hawthorn, J. \& Gerhard, O. 2016, ARA\&A, 54, 529

\bibitem[{Block {et~al.}(1994)Block, Bertin, Stockton, Grosb{\o}l, Moorwood, \&
  Peletier}]{block94}
Block, D.~L., Bertin, G., Stockton, A., {et~al.} 1994, A\&A, 288, 365

\bibitem[{Bonnarel {et~al.}(2000)Bonnarel, Fernique, Bienaym{\'e}, Egret,
  Genova, Louys, {et~al.}}]{aladin}
Bonnarel, F., Fernique, P., Bienaym{\'e}, O., {et~al.} 2000, A\&AS, 143, 33

\bibitem[{Bressan {et~al.}(2012)Bressan, Marigo, Girardi, Salasnich, Cero,
  Rubele, \& Nanni}]{bressan12}
Bressan, A., Marigo, P., Girardi, L., {et~al.} 2012, MNRAS, 427, 127

\bibitem[{Carraro(2015)}]{carraro15a}
Carraro, G. 2015, BAAS, 57, 138

\bibitem[{Carrillo {et~al.}(2018)Carrillo, Minchev, Kordopatis, Steinmetz,
  Binney, Anders, {et~al.}}]{carrillo18}
Carrillo, I., Minchev, I., Kordopatis, G., {et~al.} 2018, MNRAS, 475, 2679

\bibitem[{Drimmel(2000)}]{drimmel00}
Drimmel, R. 2000, A\&A, 358, L13

\bibitem[{Duch{\^e}ne \& Kraus(2013)}]{duchene13}
Duch{\^e}ne, G. \& Kraus, A. 2013, ARA\&A, 51, 269

\bibitem[{Englmaier \& Gerhard(1999)}]{englmaier99}
Englmaier, P. \& Gerhard, O. 1999, MNRAS, 304, 512

\bibitem[{Englmaier \& Gerhard(2006)}]{englmaier06}
---. 2006, Cel. Mech. Dyn. Astr., 94, 369

\bibitem[{Eskridge {et~al.}(2002)Eskridge, Frogel, Pogge, Quillen, Berlind,
  Davies, {et~al.}}]{eskridge02}
Eskridge, P.~B., Frogel, J.~A., Pogge, R.~W., {et~al.} 2002, ApJS, 143, 73

\bibitem[{{Gaia Collaboration} {et~al.}(2018{\natexlab{a}}){Gaia
  Collaboration}, Brown, Vallenari, Prusti, de~Bruijne, Babusiaux,
  {et~al.}}]{gaia18a}
{Gaia Collaboration}, Brown, A., Vallenari, A., {et~al.} 2018{\natexlab{a}},
  A\&A, 616, A1

\bibitem[{{Gaia Collaboration} {et~al.}(2016{\natexlab{a}}){Gaia
  Collaboration}, Brown, Vallenari, Prusti, de~Bruijne, Mignard,
  {et~al.}}]{gaia16a}
{Gaia Collaboration}, Brown, A. G.~A., Vallenari, A., {et~al.}
  2016{\natexlab{a}}, A\&A, 595, A2

\bibitem[{{Gaia Collaboration} {et~al.}(2018{\natexlab{b}}){Gaia
  Collaboration}, Katz, Antoja, Romero-G{\'o}mez, Drimmel, Reyl{\'e},
  {et~al.}}]{gaia18d}
{Gaia Collaboration}, Katz, D., Antoja, T., {et~al.} 2018{\natexlab{b}}, A\&A,
  616, A11

\bibitem[{{Gaia Collaboration} {et~al.}(2016{\natexlab{b}}){Gaia
  Collaboration}, Prusti, de~Bruijne, Brown, Vallenari, Babusiaux,
  {et~al.}}]{gaia16b}
{Gaia Collaboration}, Prusti, T., de~Bruijne, J. H.~J., {et~al.}
  2016{\natexlab{b}}, A\&A, 595, A1

\bibitem[{Gingerich(1985)}]{gingerich85}
Gingerich, O. 1985, IAUS, 106, 59

\bibitem[{Gittins \& Clarke(2004)}]{gittins04}
Gittins, D.~M. \& Clarke, C.~J. 2004, MNRAS, 349, 909

\bibitem[{Grand {et~al.}(2015)Grand, Bovy, Kawata, Hunt, Famaey, Siebert,
  {et~al.}}]{grand15}
Grand, R. J.~J., Bovy, J., Kawata, D., {et~al.} 2015, MNRAS, 453, 1867

\bibitem[{Grosb{\o}l(1993)}]{grosbol93}
Grosb{\o}l, P. 1993, PASP, 105, 651

\bibitem[{Grosb{\o}l(2016)}]{grosbol16}
---. 2016, A\&A, 585, A141

\bibitem[{Grosb{\o}l \& Patsis(1998)}]{gp98}
Grosb{\o}l, P. \& Patsis, P.~A. 1998, A\&A, 336, 840

\bibitem[{Hou \& Han(2014)}]{hou14}
Hou, L.~G. \& Han, J.~L. 2014, A\&A, 569, A125

\bibitem[{Husser {et~al.}(2013)Husser, von Berg, Dreizler, Homeier, Reiners,
  Barman, \& Hauschildt}]{husser13}
Husser, T.-O., von Berg, S.~W., Dreizler, S., {et~al.} 2013, A\&A, 553, A6

\bibitem[{Indebetouw {et~al.}(2005)Indebetouw, Mathis, Babler, Meade, Watson,
  Whitney, {et~al.}}]{indebetouw05}
Indebetouw, R., Mathis, J.~S., Babler, B.~L., {et~al.} 2005, ApJ, 619, 931

\bibitem[{Johnson \& Soderblom(1987)}]{johnson87}
Johnson, D. R.~H. \& Soderblom, D.~R. 1987, AJ, 93, 864

\bibitem[{Kawata {et~al.}(2018)Kawata, Baba, Ciuc{\v a}, Cropper, Grand, \&
  Hunt}]{kawata18}
Kawata, D., Baba, J., Ciuc{\v a}, I., {et~al.} 2018, MNRAS, 479, L108

\bibitem[{Lin \& Shu(1964)}]{lin64}
Lin, C.~C. \& Shu, F.~H. 1964, ApJ, 140, 646

\bibitem[{Lin {et~al.}(1969)Lin, Yuan, \& Shu}]{lin69}
Lin, C.~C., Yuan, C., \& Shu, F.~H. 1969, ApJ, 155, 721

\bibitem[{Liszt(1985)}]{liszt85}
Liszt, H.~S. 1985, IAUS, 106, 283

\bibitem[{Marigo {et~al.}(2017)Marigo, Girardi, Bressan, Rosenfield, Aringer,
  Chen, {et~al.}}]{marigo17}
Marigo, P., Girardi, L., Bressan, A., {et~al.} 2017, ApJ, 835, 77

\bibitem[{Mark(1976)}]{mark76}
Mark, J. W.-K. 1976, ApJ, 205, 363

\bibitem[{Mongui{\'o} {et~al.}(2015)Mongui{\'o}, Grosb{\o}l, \&
  Figueras}]{monguio15}
Mongui{\'o}, M., Grosb{\o}l, P., \& Figueras, F. 2015, A\&A, 577, A142

\bibitem[{Palacios {et~al.}(2010)Palacios, Gebran, Josselin, Martins, Plez,
  Belmas, \& L{\`e}bre}]{palacios10}
Palacios, A., Gebran, M., Josselin, E., {et~al.} 2010, A\&A, 516, A13

\bibitem[{Pasquini {et~al.}(2002)Pasquini, Avila, Blecha, Cacciari, Cayatte,
  Colless, {et~al.}}]{pasquini02}
Pasquini, L., Avila, G., Blecha, A., {et~al.} 2002, The Messenger, 110, 1

\bibitem[{Reid {et~al.}(2014)Reid, Menten, Brunthaler, Zheng, Dame, Xu,
  {et~al.}}]{reid14}
Reid, M.~J., Menten, K.~M., Brunthaler, A., {et~al.} 2014, ApJ, 783, 130

\bibitem[{Roberts(1969)}]{roberts69a}
Roberts, W.~W. 1969, ApJ, 158, 123

\bibitem[{Rodr\'{\i}guez-Merino {et~al.}(2005)Rodr\'{\i}guez-Merino, Chavez, \&
  Bertone}]{rodriguez05}
Rodr\'{\i}guez-Merino, L.~H., Chavez, M., \& Bertone, E. 2005, ApJ, 626, 411

\bibitem[{Russeil(2003)}]{russeil03}
Russeil, D. 2003, A\&A, 397, 133

\bibitem[{Saito {et~al.}(2012)Saito, Hempel, Minniti, Lucas, Rejkuba, Toledo,
  {et~al.}}]{vvv}
Saito, R.~K., Hempel, M., Minniti, D., {et~al.} 2012, A\&A, 537, A107

\bibitem[{Schlafly \& Finkbeiner(2011)}]{schlafly11}
Schlafly, E.~F. \& Finkbeiner, D.~P. 2011, ApJ, 737, 103

\bibitem[{Sch{\"o}nrich {et~al.}(2010)Sch{\"o}nrich, Binney, \&
  Dehnen}]{schonrich10}
Sch{\"o}nrich, R., Binney, J., \& Dehnen, W. 2010, MNRAS, 403, 1829

\bibitem[{Shu(1970)}]{shu70a}
Shu, F.~H. 1970, ApJ, 160, 99

\bibitem[{Siebert {et~al.}(2012)Siebert, Famaey, Binney, Burnett, Faure,
  Minchev, {et~al.}}]{siebert12}
Siebert, A., Famaey, B., Binney, J., {et~al.} 2012, MNRAS, 425, 2335

\bibitem[{Visser(1980{\natexlab{a}})}]{visser80}
Visser, H. C.~D. 1980{\natexlab{a}}, A\&A, 88, 149

\bibitem[{Visser(1980{\natexlab{b}})}]{visser80a}
---. 1980{\natexlab{b}}, A\&A, 88, 159

\bibitem[{Wu {et~al.}(2014)Wu, Sato, Reid, Moscadelli, Zhang, Xu,
  {et~al.}}]{wu14}
Wu, Y.~W., Sato, M., Reid, M.~J., {et~al.} 2014, A\&A, 566, A17

\bibitem[{Y{\'a}{\~n}ez {et~al.}(2008)Y{\'a}{\~n}ez, Norman, Martos, \&
  Hayes}]{yanez08}
Y{\'a}{\~n}ez, M.~A., Norman, M.~L., Martos, M.~A., \& Hayes, J.~C. 2008, ApJ,
  672, 207

\bibitem[{Yu \& Liu(2018)}]{yu18}
Yu, J. \& Liu, C. 2018, MNRAS, 475, 1093

\bibitem[{Yuan \& Grosb{\o}l(1981)}]{yuan81}
Yuan, C. \& Grosb{\o}l, P. 1981, ApJ, 243, 432

\end{thebibliography}

\begin{appendix}
\section{Galactic potential}
\label{apx:pot}
Although it would be preferable to compute self-gravitating test-particle
simulations, it is not currently feasible to initiate such models so that they
generate a prescribed, stable spiral pattern.  Thus, a set of test-particle
simulations were calculated in a fixed galactic potential with an imposed
spiral perturbation to estimate the velocity field of a stellar population
with ages\,$<$2\,Gyr.  The axisymmetric potential consisted of three components:
1) a bulge with a Kuzmin potential with a total mass of
$1.8\times10^9$\,M$_\sun$ and a scale length of 1.0\,pc, 2) an exponential
disk with a central surface density of 573.0\,M$_\sun$ pc$^{-2}$ and a scale
length of 2.5\,pc, and 3) a logarithmic potential for the halo with a maximum
velocity of 220\,\kms\ and a scale length of 2.0\,kpc.  This potential is
consistent with the Milky Way model of \citet{bland16} and has a rotational
speed of 244\,\kms\ at the solar radius $R_\sun = 8.36$\,kpc \citep{reid14}.

Although the length of the Galactic bar is less than 5\,kpc \citep{bland16},
it may still affect the stellar velocity field at larger distances due to its
quadruple moment.  Thus, a bar potential with a pattern speed different from
that of the spiral was included.  The spiral perturbation was not truncated in
the bar region which leads to an unrealistic potential in the very inner parts
of the Galaxy.  This is not an issue since the major objective of the models
is to estimate the velocity field between the sun and the Scutum arm, which
must be outside the bar region.

The models were populated with $2\times10^7$ test particles in the radial
range of 3.5--9.5\,kpc from the GC with a uniform distribution in azimuth and a
radial surface density variation given by the disk surface density. Due to the
symmetry of the models, there are twice as many effective   particles. The
particles were given energies corresponding to circular orbits, while their
velocity dispersion followed a Gaussian distribution with a dispersion of
10\,\kms.  The amplitude of spiral and bar potentials was increased linearly
from zero to the specified values over the first 0.6\,Gyr, after which they
were kept constant.  The orbits of the particles were integrated to an age of
1 and 2\,Gyr using a 4$\mathrm{th}$-order Runge-Kutta predictor-corrector
method with variable step ensuring a relative error of less than $10^{-6}$.
This allows the models to be dynamically relaxed, but still have ages
comparable to those of the stars measured.

The model of the Galactic potential $\Phi$ consists of the sum of three part
and depends on radius $r$, phase $\theta$, and time $t$
\begin{equation}
  \Phi(r,\theta,t) = \Phi_0(r) + \Phi_s(r,\theta) + \Phi_b(r,\theta,t)
  \label{eq:pot},
\end{equation}
where $\Phi_0$ is the axisymmetric term; $\Phi_s$ is the spiral potential, which
has no time dependence since the frame of reference is rotating with the
angular speed of $\Omega_p$; and the last term $\Phi_b$ represents the bar
potential, which may rotate with a different speed and therefore is dependent
on time.

\begin{table}[t]
  \caption[]{Distance from GC in kpc of stellar resonances in the axisymmetric
    potential as a function of pattern speed $\Omega_p$ in \kmskpc.}
\label{tab:res}
\centering
\begin{tabular}{crrrrr}
 \hline\hline
 \multicolumn{1}{c}{$\Omega_p$} & \multicolumn{1}{c}{ILR} &
 \multicolumn{1}{c}{4:1} &  \multicolumn{1}{c}{CR} &
 \multicolumn{1}{c}{-4:1} & \multicolumn{1}{c}{OLR} \\ \hline
    15.0  &   -2.06 & 10.60 & 15.67 & 20.71 & 25.82 \\
    20.0  &   -1.92 &  7.97 & 11.98 & 15.76 & 19.56 \\
    30.0  &   -2.03 &  5.07 &  8.16 & 10.79 & 13.32 \\
    40.0  &   -1.95 &  3.45 &  6.11 &  8.25 & 10.20 \\
\end{tabular}
\end{table}

The axisymmetric potential has three terms representing a bulge
(\ref{eq:blg}), an
exponential disk (\ref{eq:exp}), and a halo (\ref{eq:hal}),
\begin{eqnarray}
  \Phi_0(r) & = & -G  M_B / \sqrt(r^2 - h_B^2)  \label{eq:blg} \\
  &   & -2 \pi G \Sigma_d y (I_0(y)K_1(y) - I_1(y)K_0(y))    \label{eq:exp} \\
  &   & +0.5 V_h^2 \ln(r^2 + h_h^2) \label{eq:hal},
\end{eqnarray}
where $y = 0.5 r/h_d$ and $G$ is the gravitational constant.  The bulge was
defined by its mass $M_B = 1.8\,10^9$\,M$_\sun$ and scale length $h_B =
1.0$\,kpc. The exponential disk had a central surface density $\Sigma_d =
573$\,M$_\sun$ pc$^{-2}$ and a scale length $h_d = 2.5$\,kpc where the
functions $I$ and $K$ are the Bessel functions.  The halo had a maximum
rotational speed of $V_h = 220$\,\kms\ and the scale length $h_h = 2$\,kpc.

The spiral potential is defined by
\begin{equation}
 \Phi_s(r,\theta) = A_s r e^{-r/h_s} \cos(m(\ln(r/r_s)/\tan(i) - \theta))
  \label{eq:spr}
,\end{equation}
where $m$ is the number of arms, $i$ is the pitch angle, and  
$A_s$ is the amplitude. A scale length $h_s = 6.0$\,kpc was used to ensure
a small radial
variation of the radial force introduced by the spiral relative to the
axisymmetric force.

The bar potential was allowed to rotate with a different angular speed
$\Omega_b$ compared to that of the reference frame $\Omega_p$ and had the
form
\begin{equation}
  \Phi_b(r,\theta,t) = A_b G \cos(2(\theta_b-\theta + t(\Omega_b-\Omega_p))) /
  (r^{n_b} + h_b^{n_b})^{1/n_b}
  \label{eq:bar},
\end{equation}
where $A_b$, $\theta_b$, and $h_b$ are amplitude, initial phase,
and scale length, respectively.  The bar shape was determined by $n_b$ which
was set to 4 giving a relative fast decline of the potential with radius.
Models with $n_b = 2$ or 6 were also computed, but showed no significant
differences at the distances relevant for the current data.  The amplitude was
fixed to 10\% of the bulge mass, i.e., $M_b = 1.8\times10^8\,M_\sun$, while the
scale length $h_b$ was set to 4\,kpc. The exact analytical shape of the bar
potential is not essential for the response at distances significant outside
the bar region where the bar quadruple moment provides the main effect.

The Hamiltonian $H$ of the system rotating with the angular speed $\Omega_p$
is given as
\begin{equation}
  H(\theta,r,J,v,t) = 0.5(v^2+(J/r)^2) - J\Omega_p + \Phi(r,\theta,t),
  \label{eq:ham}
\end{equation}
where $v$ is the radial velocity and $J$ the angular momentum.  The energy is
only preserved for $A_b = 0$.

Models with pattern speeds $\Omega_p$ in the range 15--40\,\kmskpc\ were
computed, which placed the major resonances at the radii listed in
Table~\ref{tab:res}.

\section{Catalog of radial velocities}
\label{apx:data}
The catalog of the observed radial velocities of stars with S/N$>$ 10 is
available through CDS\footnote{Centre de Donn{\'e}es astronomiques de
  Strasbourg: http://cds.u-strasbg.fr} as a FITS table with the columns listed
in Table~\ref{tab:data}.  Values for $\log(g)$ and spectroscopic distances,
{\tt d-spec}, are given only for main-sequence B-stars, i.e., \lTe $>4.0$ and
$\log(g)>3.5$.

\begin{table}
\caption[]{Column specifications for FITS table with radial velocity data.}
\label{tab:data}
\begin{tabular}{llll}
 \hline\hline
 Label & Format & Unit & Remarks \\ \hline
Ident     & A10 &   - & Source identifier \\
Field     & A6  &   - & Field name \\
Fiber     & J   &   - & GIRAFFE fiber no. \\
RAdeg     & D   & deg & Right Ascension J2000 \\
DEdeg     & D   & deg & Declination J2000  \\
OBS-TIME  & D   & day & MJD of mean exposure \\
SNR       & E   &     & Signal-to-noise ratio  \\
ID-Gaia   & K   & -   & Gaia DR2 identifier \\
Gmag      & E   & mag & Gaia G magnitude \\
ID-VVV    & K   & -   & VVV identifier \\
Kmag      & E   & mag & VVV Ks MAG1AP \\
RVel      & E   & \kms\ & Barycentric radial velocity \\
RVelErr   & E   & \kms\ & Error of radial velocity \\
logTe     & E   & -   & log10 of \Te \\
logTeErr  & E   & -   & Error in log10 of \Te \\
logG      & E   & -   & log10 of surface gravity \\
Ak        & E   & mag & Extinction in Ks \\
dist      & E   & kpc & Adopted distance \\
dist-spc  & E   & kpc & Spectroscopic distance \\
start     & E   & nm  & Wavelength start \\
step      & E   & nm  & Step size of spectra \\
spec      & 3056E &   & Normalized spectrum \\
\end{tabular}
\end{table}

\section{Probabilities for models}
\label{apx:px}
The test variables $T_x$ for the observed radial velocity
distribution to be taken from the distributions computed from the
test-particle simulations are listed in the Table~\ref{tab:ps2a}. The model name
is given together with pattern speed $\Omega_p$ in \kmskpc, spiral amplitude
in \km2s2, velocity offset relative to the sun $V_o$ in \kms, phase offset
$\theta_m$ in degrees, and the test variable $T_x$.  Only models with
$T_x<9.12$ (i.e., less than that for a pure axisymmetric models), $|V_o-11.1| <
2.1$\,\kms, and $|m\theta_m| < 30\degr$ are included.

\begin{table}[t]
  \caption[]{List of test variables $T_x$ for test-particle simulations
    with a value less than that of an unperturbed model}
\label{tab:ps2a}
\centering
\begin{tabular}{lrrrrr}
 \hline\hline
 Name  &  \multicolumn{1}{c}{$\Omega_p$} &  \multicolumn{1}{c}{$A_s$} &
 \multicolumn{1}{c}{$V_o$} & \multicolumn{1}{c}{$\theta_m$} &
 \multicolumn{1}{c}{$T_x$}  \\ \hline
  {\tt s2ann} &  15 &   50 &  -9.4 &  14.1 &  8.52 \\
  {\tt s2ann} &  20 &   25 & -11.0 & 165.9 &  8.82 \\
  {\tt s2ann} &  20 &   40 & -11.0 &   2.8 &  4.39 \\
  {\tt s2ann} &  20 &   50 & -11.4 &   8.4 &  3.16 \\
  {\tt s2ann} &  20 &   60 & -11.0 &   8.4 &  2.63 \\
  {\tt s2ann} &  20 &   75 & -10.9 &  14.1 &  4.17 \\
  {\tt s2ann} &  20 &  100 & -10.7 & 177.2 &  7.49 \\
  {\tt s2ann} &  25 &   25 & -11.5 & 177.2 &  6.89 \\
  {\tt s2ann} &  25 &   50 & -11.5 &  14.1 &  4.02 \\
  {\tt s2ann} &  25 &   60 & -11.1 &  11.2 &  2.60 \\
  {\tt s2ann} &  25 &   75 & -11.1 &   5.6 &  3.09 \\
  {\tt s2ann} &  25 &  100 & -11.4 &   2.8 &  7.38 \\
  {\tt s2ann} &  25 &  125 & -10.8 &   5.6 &  6.49 \\
  {\tt s2ann} &  25 &  150 & -10.1 &   2.8 &  9.04 \\
  {\tt s2ann} &  30 &   25 & -11.7 &   2.8 &  8.44 \\
  {\tt s2ann} &  30 &   40 & -11.8 &   8.4 &  6.79 \\
  {\tt s2ann} &  30 &   50 & -11.6 &   8.4 &  5.73 \\
  {\tt s2ann} &  30 &   60 & -12.0 &  14.1 &  7.00 \\
  {\tt s2ann} &  30 &   75 & -11.4 &   5.6 &  6.05 \\
  {\tt s2ann} &  30 &  100 & -11.3 &  14.1 &  3.12 \\
  {\tt s2ann} &  30 &  125 & -11.4 &   8.4 &  3.55 \\
  {\tt s2ann} &  30 &  150 & -12.1 &   5.6 &  3.76 \\
  {\tt s2ann} &  30 &  175 & -12.2 &   8.4 &  2.96 \\
  {\tt s2cnn} &  20 &   25 & -11.1 & 177.2 &  4.20 \\
  {\tt s2cnn} &  20 &   40 & -11.5 &  11.2 &  1.81 \\
  {\tt s2cnn} &  20 &   50 & -11.5 &   5.6 &  2.79 \\
  {\tt s2cnn} &  20 &   75 & -10.5 &   2.8 &  6.91 \\
  {\tt s2cnn} &  25 &   25 & -10.9 & 165.9 &  8.57 \\
  {\tt s2cnn} &  25 &   60 & -11.6 & 168.8 &  6.41 \\
  {\tt s2cnn} &  25 &   75 & -11.5 & 168.8 &  7.25 \\
  {\tt s2cnn} &  30 &   25 & -11.6 & 165.9 &  8.56 \\
  {\tt s2cnn} &  30 &   50 & -11.6 & 171.6 &  7.73 \\
  {\tt s2cnn} &  30 &   75 & -12.3 & 165.9 &  6.65 \\
  {\tt s2cnn} &  30 &  100 & -11.7 & 168.8 &  6.89 \\
  {\tt s2dnn} &  20 &   25 & -11.3 &   0.0 &  2.37 \\
  {\tt s2dnn} &  20 &   50 & -10.7 &   0.0 &  5.86 \\
  {\tt s2dnn} &  20 &   60 & -10.1 &   5.6 &  6.91 \\
  {\tt s2dnn} &  25 &   25 & -11.0 & 165.9 &  7.01 \\
  {\tt s2dnn} &  25 &   40 & -11.6 & 171.6 &  8.02 \\
  {\tt s2dnn} &  25 &   50 & -12.2 & 174.4 &  7.07 \\
  {\tt s2dnn} &  30 &   25 & -11.8 & 165.9 &  7.30 \\
  {\tt s2dnn} &  30 &   40 & -12.3 & 165.9 &  4.77 \\
  {\tt s2dnn} &  30 &   50 & -11.7 & 174.4 &  7.13 \\
  {\tt s2enn} &  20 &   25 & -11.7 &   0.0 &  7.06 \\
  {\tt s2enn} &  20 &   40 & -10.5 &   8.4 &  3.39 \\
  {\tt s2enn} &  30 &   25 & -11.9 & 168.8 &  5.54 \\
  {\tt s2enn} &  30 &   40 & -12.7 & 165.9 &  4.45 \\
  {\tt s2enn} &  30 &   50 & -12.8 & 168.8 &  4.58 \\
  {\tt s4ann} &  15 &   50 & -11.6 &  84.4 &  8.01 \\
    \end{tabular}
\end{table}

\end{appendix}
\end{document}